\documentclass[aps,amsmath,amssymb,prl,twocolumn,showpacs]{revtex4}
\usepackage{graphicx}

\begin{document}
\title{Lifshitz Interaction between Dielectric Bodies of Arbitrary Geometry}

\author{Ramin Golestanian}
\affiliation{Institute for Advanced Studies in Basic Sciences,
Zanjan 45195-1159, Iran}

\date{\today}

\begin{abstract}
A formulation is developed for the calculation of the
electromagnetic--fluctuation forces for dielectric objects of
arbitrary geometry at small separations, as a perturbative
expansion in the dielectric contrast. The resulting Lifshitz
energy automatically takes on the form of a series expansion of
the different many-body contributions. The formulation has the
advantage that the divergent contributions can be readily
determined and subtracted off, and thus makes a convenient scheme
for realistic numerical calculations, which could be useful in
designing nano-scale mechanical devices.
\end{abstract}


\maketitle

Since the pioneering work of van der Waals in 1873---which
revealed that condensation of gases is not possible unless an
attractive interaction is at work in matter---and the explanation
of this interaction by London in 1930 in terms of the correlation
between quantum fluctuations of atoms and molecules
\cite{Israelachvili92}, we have been witnessing a surge of
interest in exploring the physical implications of such
fluctuation--induced interactions \cite{Kardar99,Bordag+01}. The
case of two parallel plates made of perfect conductors tackled by
Casimir \cite{Casimir48} and the subsequent generalization to the
case of dielectric materials by Lifshitz \cite{Lifshitz} made
important contributions to our understanding of the macroscopic
manifestations of these interactions. Relevant experimental
studies, which were ongoing alongside with the theoretical
developments \cite{Israelachvili92}, culminated recently with high
precision quantitative verifications of the Casimir force
\cite{Lamoreaux97,MR98,CAKBC2001,Bressi02}.

The advent of nanotechnology in recent years has added to the
interest in electromagnetic--fluctuation interactions as they
become dominant at nanoscale \cite{Srivastava+85}, and
quantitative knowledge of them appears to be necessary in
designing nano-machines to avoid unwanted effects such as stiction
\cite{Serry+95,Buks+01}. Moreover, one could even make use of
these interactions, {\em e.g.} the normal Casimir force between a
flat plate and a sphere \cite{CAKBC2001,chanosc} or the lateral
Casimir force between two corrugated surfaces \cite{GK,Chen+02},
in designing novel actuation schemes in microelectromechanical
systems (MEMS). It thus seems desirable to be able to calculate
the electromagnetic--fluctuation forces for a given assortment of
dielectric and metallic objects in a certain configuration.

This, however, appears to be a nontrivial task due a number of
subtleties involved. Casimir forces are known to depend in a
nontrivial way on the geometry of the objects \cite{RM99}, and the
theoretical schemes that have so far been developed to study this
effect are only applicable to perfect conductors
\cite{GK,RM99,Balian,EHGK,Jaffe,Emig}. On the other hand, Lifshitz
has pointed out that the electromagnetic--fluctuation force
between two boundaries is dominated by the dielectric properties
of the media in the frequency that is set by the separation
between them \cite{Lifshitz}. This means that when these forces
are most relevant, {\em i.e.} at length scales lower than $100$ nm
that is of the order the plasma wavelengths of most good metals,
the perfect conductor assumption in the calculation of the Casimir
force breaks down. Finally, the dependence of the divergent
contributions to the Casimir energy---that should be removed in a
carefully regularized formulation---on the geometry of the
boundaries is not well characterized, and this makes it difficult
to develop systematic numerical schemes of calculations.

\begin{figure}
\includegraphics[width=.9\columnwidth]{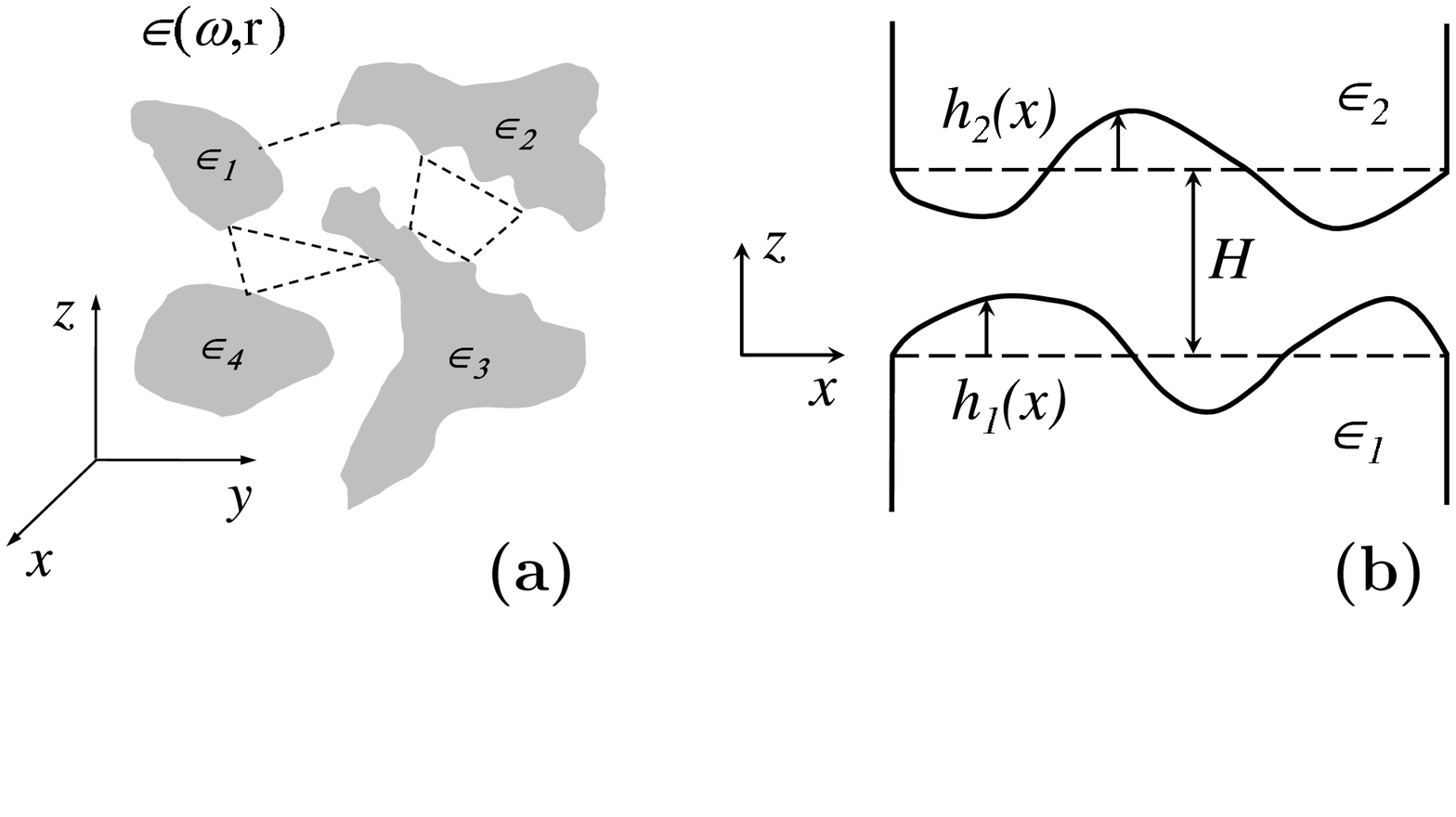}
\caption{Schematics of (a) the dielectric function profile, and
the many-body contributions to the electromagnetic--fluctuation
forces, and (b) the two semi-infinite dielectric bodies with
irregular surfaces.} \label{fig:schem}
\end{figure}

The fact that the Lifshitz interactions at small separations
effectively involve dielectric constants at relatively
high-frequencies suggests that a strategy based on expansion in
dielectric contrast could act as a useful complementary approach
to the existing formulations. Here, we have developed a path
integral formulation for calculating the Lifshitz energy for
dielectric bodies of arbitrary geometry, as an expansion in powers
of the variations in the dielectric constant in space (see Fig.
\ref{fig:schem}a). The result effectively takes on the form of an
expansion in many-body interactions with a fundamental tensorial
kernel that reflects the nature of the electromagnetic
fluctuations, with explicit expressions for all of the terms in
the series. It has the advantage that the divergent contributions
in the series are manifest and can be subtracted off
systematically, which makes it very suitable for numerical
calculations.

We start with the action integral for the electromagnetic field in
a matter with dielectric constant $\epsilon$, {\em i.e.} ${\cal
S}_{\rm m}=\frac{1}{8 \pi}\int d t d^3 {\bf x} \left[\epsilon {\bf
E}^2-{\bf B}^2\right]$, where standard expressions for the
electric field ${\bf E}=-\frac{1}{c}\partial_t {\bf A}-\nabla
\phi$ and the magnetic field ${\bf B}=\nabla \times {\bf A}$ are
assumed in terms of the potentials $\phi$ and ${\bf A}$. To
perform the quantization, we need to express the action in terms
of the potentials and break the gauge invariance by choosing a
particular gauge. Using the temporal gauge, $\phi=0$, we can write
the action integral as
\begin{math}
{\cal S}_{\rm m}=\frac{1}{8 \pi}\int \frac{d \omega}{2 \pi} \int
d^3 {\bf r} \;A_i(-\omega,{{\bf r}}) [\epsilon(\omega,{\bf r})
(\omega/c)^2 \delta_{ij}+\partial_k
\partial_k \delta_{ij}-\partial_i
\partial_j] A_j(\omega,{{\bf r}}),
\end{math}
where summation over repeated indices is assumed. In this
expression, $\epsilon(\omega,{\bf r})$ is the frequency dependent
dielectric function of the medium, which could represent any
spatial arrangement of dielectric objects of arbitrary shapes, as
depicted in Fig. \ref{fig:schem}a.

The quantization can now be performed by using the path integral
method, which is facilitated if a Wick rotation is performed in
the frequency domain. This renders the action integral
``Euclidean'' and we can find the partition function ${\cal
Z}=\int {\cal D} A_i \; \exp\left(-{\cal S}_{\rm m}^{\rm
Euclidean}/\hbar\right)$, and the Lifshitz energy
$E=-\frac{\hbar}{T} \ln {\cal Z}$, where $T$ is a long observation
time. The calculation yields
\begin{equation}
E=\hbar \int_0^\infty \frac{d \zeta}{2 \pi} \;\log \det
\left[{\cal K}_{ij}(\zeta;{\bf r},{\bf r}')\right],\label{E-1}
\end{equation}
where
\begin{math}
{\cal K}_{ij}=\left[(\zeta/c)^2 \epsilon(i \zeta,{\bf
r})\delta_{ij}+\partial_i
\partial_j-\partial_k \partial_k \delta_{ij}\right] \delta^3({\bf
r}-{\bf r}').
\end{math}
Writing $\epsilon(i \zeta,{\bf r})=1+\delta \epsilon(i \zeta,{\bf
r})$, we can decompose the kernel ${\cal K}_{ij}$ into a diagonal
part that corresponds to the empty space and a perturbation that
entails the dielectric inhomogeneity profile. In Fourier space,
this reads ${\cal K}_{ij}(\zeta;{\bf q},{\bf q}')={\cal
K}_{0,ij}(\zeta,{\bf q}) (2 \pi)^3 \delta^3({\bf q}+{\bf
q}')+\delta {\cal K}_{ij}(\zeta;{\bf q},{\bf q}')$, where ${\cal
K}_{0,ij}(\zeta,{\bf q})=(\zeta/c)^2 \delta_{ij}+q^2
\delta_{ij}-q_i q_j$ and $\delta {\cal K}_{ij}(\zeta;{\bf q},{\bf
q}')=(\zeta/c)^2 \delta_{ij} \delta \tilde{\epsilon}(i \zeta,{\bf
q}+{\bf q}')$. We can now recast the expression for the Lifshitz
energy into a perturbative series by using ${\rm tr} \ln [{\cal
K}]={\rm tr} \ln [{\cal K}_0]+\sum_{n=1}^{\infty}
\frac{(-1)^{n-1}}{n} \; {\rm tr}[({\cal K}_0^{-1} \delta {\cal
K})^n]$, where
\begin{math}\displaystyle
{\cal K}_{0,ij}^{-1}(\zeta,{\bf q})=\frac{(\zeta/c)^2
\delta_{ij}+q_i q_j}{(\zeta/c)^2 [\left(\zeta/c)^2+q^2\right]}.
\end{math}
Using these definitions, one can write the explicit form for the
trace as
\begin{eqnarray}
&&{\rm tr}[({\cal K}_0^{-1} \delta {\cal K})^n]=\int \frac{d^3
{\bf q}^{(1)}}{(2 \pi)^3} \cdots \frac{d^3 {\bf q}^{(n)}}{(2 \pi)^3} \nonumber \\
&& \hskip0.5cm \times \frac{[(\zeta/c)^2 \delta_{i_{1}
i_{2}}+q_{i_{1}}^{(1)} q_{i_{2}}^{(1)}] \cdots [(\zeta/c)^2
\delta_{i_{n} i_{1}}+q_{i_{n}}^{(n)}
q_{i_{1}}^{(n)}]}{[(\zeta/c)^2 +q^{(1)2}]
\cdots [(\zeta/c)^2 +q^{(n)2}]} \nonumber \\
&& \hskip0.5cm \times \; \delta \tilde{\epsilon}(i \zeta,-{\bf
q}^{(1)}+{\bf q}^{(2)}) \cdots \; \delta \tilde{\epsilon}(i
\zeta,-{\bf q}^{(n)}+{\bf q}^{(1)}), \label{trKKn-Fourier}
\end{eqnarray}
which involves the geometric information about the arrangement of
the dielectric objects through the Fourier transform of the
dielectric function profile. Transforming back the expression in
Eq. (\ref{trKKn-Fourier}) into real space, we find the following
series for the Lifshitz energy of any heterogeneous dielectric
medium
\begin{eqnarray}
E&=&\hbar \int_0^\infty \frac{d \zeta}{2 \pi}
\;\sum_{n=1}^{\infty}
\frac{(-1)^{n-1}}{n} \;\int d^3 {\bf r}_1 \cdots d^3 {\bf r}_n \nonumber \\
&& \times \; {\cal G}^{\zeta}_{i_{1}i_{2}}({\bf r}_1-{\bf r}_2)
\cdots \; {\cal G}^{\zeta}_{i_{n}i_{1}}({\bf r}_n-{\bf r}_1)\nonumber \\
&& \times \; \delta \epsilon(i \zeta,{\bf r}_1) \cdots \; \delta
\epsilon(i \zeta,{\bf r}_n) , \label{E-gen}
\end{eqnarray}
where the Green's function associated with the electromagnetic
fluctuations is defined as
\begin{eqnarray}
{\cal G}^{\zeta}_{ij}({\bf r})&=&\frac{(\zeta/c)^2}{4 \pi}
\frac{{\rm e}^{-\zeta r/c}}{r} \left[\delta_{ij}
\left(1+\frac{c}{\zeta r}+\frac{c^2}{\zeta^2 r^2}\right)\right.\nonumber\\
&&\left.-\frac{r_i r_j}{r^2} \left(1+3 \frac{c}{\zeta r}+3
\frac{c^2}{\zeta^2 r^2}\right)\right]+\frac{1}{3} \delta_{ij}
\delta^3({\bf r}). \label{Gij-1}
\end{eqnarray}
This result has a number of interesting characteristics. First, it
appears that an expansion in powers of $\delta \epsilon$
automatically turns into a summation of integrated contributions
of $n$-body interactions. Moreover, all of the $n$-body
interaction terms have simple expressions in terms of a single
fundamental kernel ${\cal G}^{\zeta}_{ij}({\bf r})$ that mediates
the two-body part of the interaction. This kernel is tensorial,
and has the structure of the electric field of a radiating dipole
in imaginary frequency \cite{Jackson}. It has, in fact, been
introduced some time ago in connection with van der Waals
interactions \cite{Green}. Finally, the result has a closed form
expression for the Lifshitz energy for any geometrical arrangement
of dielectric bodies in terms of quadratures.

\begin{figure}
\includegraphics[width=.7\columnwidth]{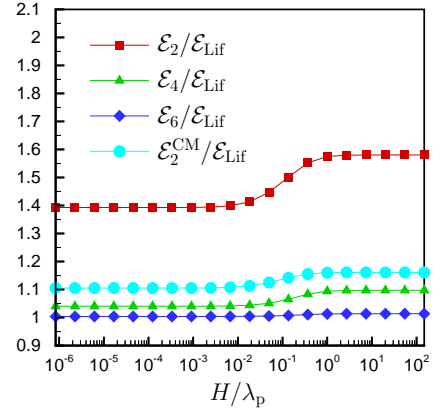}
\caption{Comparison of the expansion in Eq. (\ref{E-gen}) up to
the second, fourth, and sixth order, as well as the
Clausius-Mossotti approximation, with the exact Lifshitz result
for a pair of parallel semi-infinite dielectrics, as a function of
their separation $H$ in units of the plasma wavelength
$\lambda_{\rm p}$. The dielectric constant of both objects is
chosen as $\epsilon(i \zeta)=1+\omega_{\rm
p}^2/(\omega_0^2+\zeta^2)$ with $\omega_0^2=2 \;\omega_{\rm
p}^2$.} \label{fig:e246CM}
\end{figure}

To examine the convergence property of the above series, we can
take a specific example for which the exact result is known, and
compare it with what we find from the first few terms in Eq.
(\ref{E-gen}). We consider the case of two identical semi-infinite
dielectric objects that are placed parallel to each other at a
separation $H$, for which the exact expression for energy per unit
area is known to be
\begin{math}
{\cal E}_{\rm Lif}=\frac{\hbar}{4 \pi^2 c^2} \int_0^{\infty} d
\zeta \; \zeta^2 \int_1^{\infty} d p \; p
\left\{\ln\left[1-\frac{(s-p)^2}{(s+p)^2} {\rm e}^{-2 p \zeta
H/c}\right]\right.+\left.\ln\left[1-\frac{(s-p \epsilon)^2}{(s+p
\epsilon)^2} {\rm e}^{-2 p \zeta H/c}\right]\right\},
\end{math}
where $s=\sqrt{\epsilon-1+p^2}$ \cite{Lifshitz}. To calculate the
energies, we have assumed a simple form
\begin{math}\displaystyle
\epsilon(i \zeta)=1+\frac{\omega_{\rm p}^2}{\omega_0^2+\gamma
\zeta+\zeta^2}
\end{math}
for the dielectric constant, where $\omega_{\rm p}$ represents the
plasma frequency, from which the plasma wavelength $\lambda_{\rm
p}=2 \pi c/\omega_{\rm p}$ can be extracted. In Fig.
\ref{fig:e246CM}, the ratio between the energy as calculated from
Eq. (\ref{E-gen}) up to the second, fourth, and sixth order and
the exact Lifshitz result is shown as a function of the separation
in units of the plasma wavelength for $\omega_0^2=2 \;\omega_{\rm
p}^2$ and $\gamma=0$, which corresponds to a dielectric with
$\epsilon(0)=1.5$. The results clearly show a crossover between
two asymptotic regimes near $H \approx \lambda_{\rm p}/(2 \pi)$
\cite{note1}. The series appears to be rapidly convergent so long
as $\delta \epsilon(0)<1$, and the convergence is considerably
more efficient for $H < \lambda_{\rm p}/(2 \pi)$ [as compared to
$H > \lambda_{\rm p}/(2 \pi)$], especially for $\delta \epsilon(0)
\lesssim 1$. The damping constant $\gamma$ appears to have very
little effect on the Lifshitz energy for dielectrics ({\em i.e.}
when $\omega_0 \neq 0$). The results presented in Fig.
\ref{fig:e246CM}, for example, do not change appreciably for
nonvanishing damping constants of up to $\gamma=0.1 \; \omega_{\rm
p}$.

It is quite well known that any calculation of the Casimir or
Lifshitz energy encounters a variety of divergent contributions
that are very difficult to characterize. For example, it is not
clear how these divergent terms depend on the geometry of the
system, so that they could be identified in a simple geometry and
done away with in a systematic way for slightly deformed
boundaries, in the calculations that involve perturbation in the
geometry of the objects \cite{GK,EHGK}. While the present
formulation also suffers from this deficiency, in the sense that
the expression in Eq. (\ref{E-gen}) involves divergent
contributions, the fact that the expressions for the various terms
in the series are known {\em a priori} allows for systematic
identification of these contributions and thus their systematic
cancelation. For example, one can show that putting $1$ instead of
the fraction in the second line of Eq. (\ref{trKKn-Fourier})
yields a singular contribution of the form
\begin{math}
E_{\rm sing.}=\hbar \int_0^\infty \frac{d \zeta}{2 \pi} \int d^3
{\bf r} \; \ln \left[\epsilon(i \zeta,{\bf r})\right] \int
\frac{d^3 {\bf q}}{(2 \pi)^3},
\end{math}
which could be subtracted off systematically (i.e. at every order
$n$ in the series expansion) for any geometry and dielectric
configuration.

Let us now focus our attention on the specific arrangement shown
in Fig. \ref{fig:schem}b, where two semi-infinite dielectric
bodies with irregularly shaped boundaries are placed nearly
parallel to each other at a mean separation $H$. We can write down
the dielectric function profile as
\begin{equation}
\epsilon(i \zeta,{\bf r})=\left\{\begin{array}{ll}
\epsilon_2(i \zeta), & \; H+h_2({\bf x}) \leq z < +\infty,  \\ \\
1, & \; h_1({\bf x}) < z < H+h_2({\bf x}),  \\ \\
\epsilon_1(i \zeta), & \; -\infty < z \leq h_1({\bf x}),
\end{array} \right. \label{epsilon-profile}
\end{equation}
and the corresponding Fourier transform as
\begin{math} 
\delta \tilde{\epsilon}(i \zeta,{\bf q})=\frac{i}{q_z} \int d^2
{\bf x} \; {\rm e}^{i {\bf q}_{\perp} \cdot {\bf x}} [\delta
\epsilon_2 \; {\rm e}^{i q_z \left[H+h_2({\bf x})\right]}-\delta
\epsilon_1 \; {\rm e}^{i q_z h_1({\bf x})}].
\end{math}
When the separation of the surfaces is smaller than the plasma
wavelengths of the two dielectric media, and $\delta \epsilon_1$
and $\delta \epsilon_2$ are small compared to unity, the leading
contribution in Eq. (\ref{E-gen}) comes from the second order
term. Putting in the dielectric function profile, we find
\begin{eqnarray}
E_2&=&-\frac{\hbar}{128 \pi^3 c^4}\int_0^\infty d \zeta \; \zeta^4
\; \delta \epsilon_1(i \zeta) \delta \epsilon_2(i \zeta) \nonumber
\\
&\times& \int d^2 {\bf x} d^2 {\bf x}'\;{\cal
W}\left(\frac{\zeta}{c}[{\bf x}-{\bf
x}'],\frac{\zeta}{c}[H+h_2({\bf x})-h_1({\bf
x}')]\right),\nonumber \\
\label{E-2-1}
\end{eqnarray}
where
\begin{eqnarray}
{\cal W}({\bf y},h)=8 \;\Gamma\left(0,2
\sqrt{y^2+h^2}\right)+\int_{1}^{\infty} \frac{d s}{s^{3/2}}\;
{\rm e}^{-2 \sqrt{y^2+h^2 s}} \nonumber \\
\times\left[\frac{3}{(y^2+h^2 s)^2}+\frac{6}{(y^2+h^2
s)^{3/2}} +\frac{4 \;(1-h^2 s)}{(y^2+h^2 s)}\right],\nonumber \\
\label{Mxx'}
\end{eqnarray}
with the incomplete gamma function defined as
$\Gamma(a,z)=\int_z^\infty d t \;t^{a-1}\;{\rm e}^{-t}$. Note that
at this order, the Lifshitz energy is pairwise additive. In the
above result, we have kept the frequency dependence of the
dielectric functions as well as the geometry of the boundaries
arbitrary for generality of the presentation \cite{Barton}. The
expression in Eq. (\ref{E-2-1}) can be considerably simplified for
$h_1({\bf x})=0$:
\begin{equation}
\left.E_2\right|_{h_1({\bf x})=0}=\int d^2 {\bf x}\; {\cal
E}_2\left(H+h_2({\bf x})\right),\label{E-2-h1=0}
\end{equation}
in terms of the original Lifshitz result for the energy per unit
area of flat boundaries \cite{Lifshitz}
\begin{equation}
{\cal E}_2(H)=-\frac{\hbar}{64 \pi^2 c^2}\int_0^\infty d \zeta \;
\zeta^2 \; \delta \epsilon_1(i \zeta) \delta \epsilon_2(i \zeta)
{\cal L}\left(\frac{\zeta H}{c}\right),\label{E-2-h1=0-2}
\end{equation}
with
\begin{math}
{\cal L}(u)=4 (u^2-1) E_1(2 u)+\frac{{\rm e}^{-2 u}}{u^2} (1+2
u+u^2-2 u^3),
\end{math}
and the exponential integral function defined as
$E_n(z)=\int_1^\infty d t \;{\rm e}^{-z t}/t^n$. This
simplification is a general feature of pairwise additive
interactions, as shown in Ref. \cite{EHGK}.

The pairwise summation approximation is widely used in the
literature not only for dielectric materials at close separations
but also for perfect conductors at arbitrary separations. We can
use the present formulation to shed some light on the nature of
this approximation, and make assessment on what is not included.
Going back to Eq. (\ref{Gij-1}), we note that the Green's function
${\cal G}^{\zeta}_{ij}({\bf r})$ is decomposed into a long-ranged
kernel, and a ``contact'' part in the form of $\frac{1}{3}
\delta_{ij} \delta^3({\bf r})$. A systematic pairwise summation
approximation then amounts to keeping two of the ${\cal
G}^{\zeta}_{ij}$'s in its full form and approximating the rest of
them by their contact-contribution, in each term of the series in
Eq. (\ref{E-gen}). Considering the different ways of doing this
and keeping track of the corresponding combinatorial prefactors,
one can then sum up the entire series and find a closed form
expression for the Lifshitz energy between objects of arbitrary
shapes in the pairwise summation approximation. The result will be
identical to Eq. (\ref{E-2-1}) except for the replacement
\begin{math}
\delta \epsilon_i \to \frac{\delta \epsilon_i}{1+(\delta
\epsilon_i/3)}=3 \left(\frac{\epsilon_i-1}{\epsilon_i+2}\right).
\end{math}
The combination reminds us of the Clausius-Mossotti equation for
molecular polarizability, which is well known to be valid only for
dilute materials such as gases \cite{Jackson}. This approximation
allows us to calculate the energy for perfect conductors, and we
can use the example of two flat and parallel perfect conductors,
for which the exact result is known to be ${\cal E}^c_{\rm
exact}=-\frac{\pi^2}{720} \frac{\hbar c}{H^3}$, to assess its
validity. Letting $\epsilon_1,\epsilon_2 \to \infty$, we find
${\cal E}^c_{\rm PWS}=-\frac{69}{640 \pi^2} \frac{\hbar c}{H^3}$,
which compares to the exact result as
\begin{math} 
\frac{{\cal E}^c_{\rm PWS}}{{\cal E}^c_{\rm exact}}=\frac{621}{8
\pi^4} \simeq 0.797.
\end{math}
While this ratio varies with geometry, it gives us an estimate of
typical errors that are involved in the calculations based on
pairwise summation approximation. We also note that any attempt in
going beyond this approximation should include the tensorial
structure involved in Eq. (\ref{E-gen}), and for example, an {\em
ad hoc} augmentation by introduction of scalar three-body
interactions {\em etc.} will not be justified in light of our
present scheme. The Clausius-Mossotti approximation appears to
give comparatively better results for the case of dielectrics, as
can be seen from the example in Fig. \ref{fig:e246CM}.

The formulation presented here could also be applied to the case
of magnetic materials \cite{mag}. For a medium that is described
by the dielectric function profile $\epsilon(\omega,{\bf r})$ and
the magnetic permeability profile $\mu(\omega,{\bf r})$, we should
replace the kernel in Eq. (\ref{E-1}) by
\begin{math}
{\cal K}_{ij}(\zeta;{\bf r},{\bf r}')=[(\zeta/c)^2 \epsilon(i
\zeta,{\bf r})\delta_{ij}+\partial_i \frac{1}{\mu(i \zeta,{\bf
r})}\partial_j-\partial_k \frac{1}{\mu(i \zeta,{\bf r})}
\partial_k \delta_{ij}] \delta^3({\bf r}-{\bf r}'),
\end{math}
and the rest of the procedure follows closely. Generalization to
the case of finite temperatures by discretizing the frequency is
also straightforward.

In conclusion, we have presented a path integral formulation for
the calculation of the Lifshitz energy for dielectric materials of
arbitrary shape, as a series expansion in the dielectric contrast.
The expansion converges very rapidly for dielectric objects that
are at separations considerably smaller than their corresponding
plasma wavelengths, and is expected to work perfectly for surfaces
at nanometric separations. The results presented here could be
applicable for the calculation of electromagnetic--fluctuation
forces that are involved in nano-mechanical devices.


It is a great pleasure to acknowledge fruitful discussions with
M.A. Charsooghi, T. Emig, A. Hanke, M. Kardar, R. Matloob, M.F.
Miri, and F. Mohammad-Rafiee.


\begin{thebibliography}{99}

\bibitem{Israelachvili92}
J.N. Israelachvili, {\em Intermolecular and Surface Forces}
(Academic, London, 1992).

\bibitem{Kardar99}
M. Kardar and R. Golestanian, Rev. Mod. Phys.
{\bf 71}, 1233 (1999).

\bibitem{Bordag+01}
M. Bordag, U. Mohideen, V.M. Mostepanenko,
Phys. Rep. {\bf 353}, 1 (2001).

\bibitem{Casimir48}
H.B.G. Casimir, Proc. K. Ned. Akad. Wet. {\bf 51}, 793 (1948).

\bibitem{Lifshitz}
E.M. Lifshitz, Sov. Phys. JETP {\bf 2}, 73 (1956); I.E.
Dzyaloshinskii, E.M. Lifshitz, and L.P. Pitaevskii, Adv. Phys.
{\bf 10}, 165 (1961).

\bibitem{Lamoreaux97}
S.K. Lamoreaux, Phys. Rev. Lett. {\bf 78}, 5 (1997); {\bf 81},
5475(E) (1998); Phys. Rev. A {\bf 59}, R3149 (1999).

\bibitem{MR98}
U. Mohideen and A. Roy, Phys. Rev. Lett. {\bf 81}, 4549 (1998);
B.W.~Harris, F.~Chen, and U.~Mohideen, Phys. Rev. A {\bf 62},
052109 (2000).

\bibitem{CAKBC2001}
H.B. Chan, V.A. Aksyuk, R.N. Kleiman, D.J. Bishop, and F. Capasso,
Science {\bf 291}, 1941 (2001).

\bibitem{Bressi02}
G. Bressi, G. Carugno, R. Onofrio, and
G. Ruoso, Phys. Rev. Lett. {\bf 88}, 041804 (2002).

\bibitem{Srivastava+85} Y.~Srivastava, A. Widom, and M.H. Friedman,
Phys. Rev. Lett. {\bf 55}, 2246 (1985); M.A. Stroscio, Phys. Rev.
Lett. {\bf 56}, 2107 (1986).

\bibitem{Serry+95} F.M. Serry, D. Walliser, and G.J. Maclay,
J. Microelectromech. Syst. {\bf 4}, 193 (1995); J. Appl. Phys.
{\bf 84}, 2501 (1998).

\bibitem{Buks+01} E.~Buks and M.L. Roukes, Phys. Rev. B {\bf 63},
033402 (2001); Nature {\bf 419}, 119 (2002).

\bibitem{chanosc}
H.B. Chan, V.A. Aksyuk, R.N. Kleiman,
D.J. Bishop, and F. Capasso, Phys. Rev. Lett. {\bf 87}, 211801
(2001).

\bibitem{GK}
R. Golestanian and M. Kardar, Phys. Rev. Lett. {\bf 78}, 3421
(1997); Phys. Rev. A {\bf 58}, 1713 (1998).

\bibitem{Chen+02} F. Chen, U. Mohideen, G.L. Klimchitskaya, and
V.M. Mostepanenko, Phys. Rev. Lett. {\bf 88}, 101801 (2002); Phys.
Rev. A {\bf 66}, 032113 (2002).

\bibitem{RM99}
A. Roy and U. Mohideen, Phys. Rev. Lett. {\bf 82},
4380 (1999).

\bibitem{Balian}
R. Balian and B. Duplantier,
Ann. Phys. (New York) {\bf 104}, 300 (1977); {\bf 112}, 165
(1978).

\bibitem{EHGK}
T. Emig, A. Hanke, R. Golestanian, and M. Kardar, Phys. Rev. Lett.
{\bf 87}, 260402 (2001); Phys. Rev. A {\bf 67}, 022114 (2003).

\bibitem{Jaffe}
R.L. Jaffe and A. Scardicchio, Phys. Rev. Lett. {\bf 92}, 070402
(2004).

\bibitem{Emig}
T. Emig, Europhys. Lett. {\bf 62}, 466 (2003); R. B\"uscher and T.
Emig, Phys. Rev. Lett. {\bf 94}, 133901 (2005).

\bibitem{Jackson}
J.D. Jackson, {\em Classical Electrodynamics} (Wiley, New York,
1999).

\bibitem{Green}
I. Brevik and J.S. H{\o}ye, Physica A (Amsterdam) {\bf 153}, 420
(1988).

\bibitem{note1}
Note that the last few data points in Fig. \ref{fig:e246CM}
corresponding to large values of $H/\lambda_{\rm p}$ would almost
certainly lie in the thermal regime where the sum over the
frequency should be discretized \cite{Lifshitz}. These points are
only presented to help demonstrate the crossover behavior in a
convincing way.

\bibitem{Barton}
For a related work see: G. Barton, J. Phys. A {\bf 34}, 4083
(2001).

\bibitem{mag}
T.H. Boyer, Phys. Rev. A {\bf 9}, 2078 (1974); V. Hushwater, Am.
J. Phys. {\bf 65}, 381 (1997); O. Kenneth, I. Klich, A. Mann, and
M. Revzen, Phys. Rev. Lett. {\bf 89}, 033001 (2002).

\end{thebibliography}
\end{document}